\documentstyle[12pt,aasms4]{article}

\received{1997 December 29}
\accepted{1998 March 16}

\slugcomment{
Tp be appeared in The Astonomical Journal, 1998 July issue
}

\begin{document}

\newcommand {\etal}{{\it et al.}}
\newcommand {\HI}{H~{\small I}}
\newcommand {\HII}{H~{\small II}}
\newcommand {\HA}{H$\alpha$}
\newcommand {\HB}{H$\beta$}
\newcommand {\NII}{[N~{\small II}]}
\newcommand {\OIII}{[O~{\small III}]}
\newcommand {\degree}{$^{\circ}$}
\newcommand {\Msolar}{$M_{\odot}$}
\newcommand {\Lsolar}{$L_{\odot}$}
\newcommand {\LFIRLB}{$L_{FIR}/L_{B}$}
\newcommand {\fFIRfB}{$f_{FIR}/f_{B}$}
\newcommand {\MHILB}{$M_{\rm H I}/L_{B}$}

\title{
H-Alpha Velocity Fields of H~II Regions
in Nearby Dwarf Irregular Galaxies
\altaffilmark{1}
}

\altaffiltext{1}{
Based on observations made at Okayama Astrophysical Observatory (OAO).
OAO is a branch of the National Astronomical Observatory of Japan,
an inter-university research institute operated by
the Ministry of Education, Science, and Culture, Japan.}

\author{
Akihiko Tomita
}
\affil{
Department of Earth and Astronomical Sciences,
Faculty of Education,
\\ Wakayama University,
930 Sakae-Dani, Wakayama 640-8510, Japan
\\
Electronic mail: atomita@center.wakayama-u.ac.jp
}

\author{
Kouji Ohta,
Kouichiro Nakanishi,
Tsutomu T. Takeuchi \altaffilmark{2}, and
Mamoru Sait\={o}
}
\affil{
Department of Astronomy,
Faculty of Science, 
Kyoto University
\\
Sakyo-ku, Kyoto 606-8502, Japan
\\
Electronic mail:
ohta, nakanisi, takeuchi, saitom@kusastro.kyoto-u.ac.jp
}

\altaffiltext{2}{
Research Fellow of the Japan Society for the Promotion of Science.
} 

\begin{abstract}

We present \HA\ velocity fields of thirteen giant \HII\ regions
in four nearby dwarf irregular galaxies,
NGC~2366, Holmberg~II, IC~2574, and WLM.
We classify the velocity features
as well as the morphologies of the \HII\ regions.
The \HA\ velocity features are divided into three categories;
three \HII\ regions show chaotic feature with a typical scale of variation of
a few 100~pc in size and a few 10~km~s$^{-1}$ in velocity,
one shows expanding-bubble feature,
and the remaining nine have calm velocity fields.
There is a correlation between the \HA\ velocity feature and
the morphology of \HII\ regions.
We measured bulk motion of the \HII\ regions
relative to the ambient \HI\ velocity
for the \HII\ regions with calm velocity fields
and found a typical velocity difference of about 5~km~s$^{-1}$.
We discuss a model for origin of star-forming regions
based on the presence of the velocity difference
between \HA\ and \HI\ gas as well as the \HI\ characteristics.

\end{abstract}

\keywords{
galaxies: dwarf ---
galaxies: irregular ---
galaxies: ISM ---
galaxies: individual (NGC~2366, Holmberg~II, IC~2574, WLM)
}

\section{Introduction}

Since dwarf irregular galaxies are slow and nearly rigid rotators,
they are suitable for studying kinematical and morphological properties of
star-forming regions and interaction with surrounding interstellar medium (ISM).
Some dwarf irregulars have many \HI\ holes
(e.g., \cite{puche92} 1992);
relation between the \HI\ holes and the star formation activity was
discussed but it is not yet clear.
\cite{roy92} (1992) claimed that an extremely high-velocity exploding component
is commonly associated with \HII\ regions in dwarf irregular galaxies,
but nature of the high-velocity component has not been understood.
Star formation activities of some dwarf irregular galaxies,
even of isolated ones,
are as high as those of spiral galaxies (\cite{hunter86} 1986),
though the dwarfs do not have spiral arms.
\cite{saito92} (1992) proposed a model that
collision of \HI\ clouds from an extended \HI\ envelope
with a dense \HI\ disk could be a star formation trigger
based on the observations of the \HA\ velocity field in IC~10.
\cite{tomita93} (1993, 1994) presented the \HA\ velocity fields
in other four dwarf irregular galaxies
and showed results which are consistent with
the model by \cite{saito92} (1992).
However,
the number of samples is still small and
it is necessary to obtain more data to examine this model.

In this paper we present \HA\ velocity fields
in \HII\ regions of another four nearby dwarf irregular galaxies
and study the connection between the \HA\ velocity field
and the characteristics of the \HII\ regions.
In Sec.~2,
we describe the sample galaxies,
observations,
and data reduction.
We show the resultant position-velocity diagrams
of the \HA\ emission in Sec.~3.
Discussions are given in Sec.~4.

\section{Observations}

\subsection{Sample}

We observed four galaxies,
NGC~2366, Holmberg~II, IC~2574, and WLM
(the Wolf-Lundmark-Melotte system).
They were selected from nearby,
within about 3~Mpc,
well-studied dwarf irregular galaxies
at the location observable at the Okayama Astrophysical Observatory in Japan.
We also preferred a small apparent inclination of the galaxy
in order to avoid overlapping the \HII\ regions along the line of sight.

Table~1 lists basic characteristics of the four observed galaxies.
The second column tabulates the distance.
NGC~2366, Holmberg~II, and IC~2574 are
members of the M~81 group,
and the distances are about 3~Mpc.
WLM is a member of the Local Group,
and the distance is about 1~Mpc.
The third column tabulates physical size of diameter of the galaxy.
Three galaxies in the M~81 group have diameters of about 8 to 12~kpc
and WLM has a smallest diameter of about 3~kpc.
The fourth column tabulates the apparent inclination,
where 90\degree\ refers to edge-on.
The sixth column tabulates far-infrared (FIR) luminosity,
which is an indicator of the present star formation rate of the galaxy.
The seventh column tabulates the $B$-band luminosity,
which is a measure of the luminous mass of the galaxy.
The eighth column tabulates the dynamical total mass of the galaxy
inferred from the \HI\ observations.
The last column tabulates the \HI\ mass.

We use the \HI\ velocity field data in literature
to compare with our \HA\ velocity fields.
The high-resolution synthetic \HI\ observations were made
for NGC~2366 by \cite{wevers86} (1986) and \cite{braun95} (1995),
for Holmberg~II by \cite{puche92} (1992),
and for IC~2574 by \cite{martimbeau94} (1994).
For WLM there is no synthetic observation,
therefore,
we do not have the \HI\ velocity field map with
a high spatial resolution
suitable for comparison with our \HA\ data.
We searched for the velocity data of CO molecular emission in literature,
and found only one report for NGC~2366 by \cite{huntersage93} (1993)
with no detection of the emission.

\subsection{Observations and Reduction}

A long-slit \HA\ spectroscopy was carried out in 1994
at the Okayama Astrophysical Observatory (OAO) in Japan
using the 1.88-m reflector with the Spectronebulagraph
(an automatic slit scanning system,
see \cite{kosugi95} 1995).
We observed giant \HII\ regions in four galaxies;
for NGC~2366 most part of the giant \HII\ complexes
at the southern part of the galaxy
and some positions at the northern part of the galaxy,
for Holmberg~II about a half of the \HII\ regions
which concentrate at the central part of the galaxy,
for IC~2574 about a half of the most prominent \HII\ complexes
at the northwestern part of the galaxy,
and for WLM most of the \HII\ regions.
The observed slit positions are shown on the \HA\ map of Fig.~1 and
the observed \HII\ regions are summarized in Table~2.
Detailed description about individual \HII\ region is given in Sec.~3.2.

A spectrograph with a grating of 1200 grooves mm$^{-1}$
blazed at 7500 \AA\ and a Photometrics CCD with 512 $\times$ 512 pixels
(pixel size of 20~$\mu$m $\times$ 20~$\mu$m) was
equipped at the Cassegrain focus.
The dispersion was 0.7~\AA\ pixel$^{-1}$,
or 30~km~s$^{-1}$~pixel$^{-1}$ at the \HA\ line.
The slit was 5.$'$0 long and 1.$''$8 wide
and the instrumental broadening corresponded to about
1.3~\AA\ in FWHM.
The slit length was longer than the size of each \HA\ emitting region
and by using the spectra of the sky we could make a sky subtraction well.
A CCD pixel corresponds to 0.$''$75.
A typical seeing size was 2$''$ to 3$''$.
A log of the observations is listed in Table~3.

The data reduction and analysis were performed with the IRAF
in the usual manner
(IRAF is the software developed in
National Optical Astronomy Observatories).
We binned the spectra in four adjacent CCD pixels (3.$''$0)
along the slit length at all of positions,
except for the positions where the spectra have low signal-to-noise ratios,
where we binned in eight pixels (6.$''$0).
Since most of the observed \HA\ emission-line profiles are symmetric
and nearly Gaussian,
we fitted a single Gaussian to the observed line
to measure the central wavelength and FWHM of the line.
The error in determining the line center is about 3~km~s$^{-1}$
(corresponding to one tenth of the pixel size)
to 20~km~s$^{-1}$ depending on the signal-to-noise ratios.
The error of the measured FWHM is similar to that of the measured
central wavelength.

\section{Results}

\subsection{Position-Velocity Diagram}

We present all of our results in the form of
position-velocity diagrams in Figs.~2 -- 5.
The abscissa indicates the position along the slit,
and the width of the abscissa
corresponds to the line length given in Fig.~1.
The origin on the slit position,
shown as 0$''$ on the abscissa,
is listed in Table~4.
The ordinates of diagrams in Fig.~2 indicate
the heliocentric radial velocity of line center
and the FWHM of line.
The scale for the the central velocity is shown on the left-side ordinate
and that for the FWHM is shown on the right-side ordinate.
The filled circles indicate
the central velocities of line profiles
and the open circles indicate the FWHM of the lines.
The horizontal error bar on a filled circle represents
the binning size and
the vertical error bar indicates an error
in the measurement of the central velocity of a line.
Within each panel,
relative \HA\ intensities are propotional to the signal-to-noise ratios and
these are almost inversely propotional to sizes of the vertical error bars.
We corrected the instrumental broadening to obtain the FWHM.
The FWHMs shown are, thus,
due to the thermal broadening
which is estimated to be about 10~km~s$^{-1}$
corresponding to $10^{4}$~K
and the motion of the ionized gas
integrated in the line of sight.
The dotted line in each diagram shows the 0~km~s$^{-1}$-level of the FWHM;
the correction for the instrumental broadening could make
negative values of FWHM for some lines with low signal-to-noise ratios.
The thick curve indicates the \HI\ velocity taken from literature.
The cross in the upper-left corner of the panel
shows the accuracy of the \HI\ observations.
The full length of the horizontal line of the cross
indicates the FWHM of the beam for the \HI\ isovelocity map.
The half length of the vertical line of the cross
indicates the channel spacing of the \HI\ observations.

\subsection{Results of Individual Galaxies}

\subsubsection{NGC 2366}

The resultant sixteen position-velocity diagrams are shown in Fig.~2,
where the length of abscissa corresponds to 160$''$ or
2.7~kpc at NGC~2366.
The \HII\ regions are concentrated at the southern part of the galaxy,
and the most prominent \HII\ complex has another name of NGC~2363.
There is the second prominent \HII\ complex at the east of NGC~2363;
following \cite{arsenault86} (1986)
we call this \HII\ region NGC~2366-III.

In NGC~2363 the \HA\ velocity field is calm at the central regions
where the intensities of the \HA\ emission are most powerful;
in the diagrams the central regions are located on
$20''$ to $-10''$ at the slit position IA,
$150''$ to $130''$ at the slit positions SE, SF, SG, and SH.
The \HA\ velocity is almost the same as the \HI\ velocity,
and the line width is almost constant at 20~km~s$^{-1}$.
The FWHM of about 20~km~s$^{-1}$ and the calm velocity field
at the central regions of NGC~2363 are consistent with the measurement
by \cite{arsenault86} (1986) who obtained
the velocity dispersion of the integrated \HA\ profile of NGC~2363
(denoted as NGC~2366~I in their paper) to be 23~km~s$^{-1}$.
\cite{terlevich81} (1981) showed that NGC~2363 is
on the size-velocity dispersion relation of the extragalactic \HII\ regions
which is expected for self-gravitating systems.

Unlike at the central regions,
the \HA\ velocity field is curved
at the outer regions in NGC~2363;
see the slit positions IC, ID, SA, SB, SC, and SD.
The typical scale of the variation of velocity
is $20''$ to $30''$ corresponding to about 200~pc in size,
and 10 to 30~km~s$^{-1}$ in velocity.
An expanding shell-like velocity feature is seen at the position of
$180''$ to $150''$ on the slit SB,
east side of NGC~2363.
By comparing with the \HA\ image of Fig.~1 (a),
location of the shell-like velocity feature corresponds to
\HA\ filaments.

The slit positions of IA to ID were located
crossing the northwestern part of NGC~2363.
From the center to outer region of NGC~2363 (from IA to ID),
the \HA\ velocity rises from 85 to 110~km~s$^{-1}$,
drifting away from the \HI\ velocity.
The line width is almost constant at 20~km~s$^{-1}$
except for the regions with poor signal-to-noise ratios.
The rise of the \HA\ velocity is also shown in diagrams
of the slit positions SA to SC
as an upturn velocity gradient at around $150''$ to $130''$.
The variation of the position-velocity diagrams is consistent with
the \OIII\ velocity field given by \cite{roy91} (1991)'s Fig.~7,
in which a high velocity ridge with an altitude of about 10~km~s$^{-1}$
appears in the northwest direction from the core of NGC~2363.

The \HA\ velocity of the second prominent \HII\ region,
NGC~2366-III,
is shown at around the position of 90$''$ to 50$''$ on the slits SA to SH.
The velocity field is calm in general,
though the \HA\ velocity is curved at the slit positions SF to SH.
The \HA\ velocities at the intense \HA\ regions are blue-shifted by
several to 10~km~s$^{-1}$ than the \HI\ velocity.

The \HII\ regions at the northern part of the galaxy are
one to two orders of magnitude less intense in the \HA\ light
compared with the southern two prominent ones.
The \HA\ velocity field shown on the slit positions NA to ND
is not so curved and nearly the same as the \HI\
velocity,
though the signal-to-noise ratio is poor.
Only two regions,
at the position of 170$''$ to 190$''$ on the slit NA and
at the position of 180$''$ to 200$''$ on the slit ND,
have sufficient signal-to-noise ratios.
Each of these two regions corresponds to small circular \HII\ regions
(see Fig.~1(a));
hereafter we call them the \HII\ region NGC~2366~NA and ND,
respectively;
see Fig.~1 (a) and Table~2.

\subsubsection{High-Velocity Expanding Component in NGC~2363}

\cite{roy91} (1991) showed that at the central region of NGC~2363
the profiles of the emission lines split
corresponding to an expanding velocity of 45~km~s$^{-1}$,
and that whole NGC~2363 has a high-velocity expanding component
with a velocity width of about 1000~km~s$^{-1}$.
\cite{roy92} (1992) and \cite{gonzalez94} (1994)
reported confirmations of the high-velocity expanding component,
and \cite{roy92} (1992) claimed that any models could not explain
the large expanding velocity.
\cite{hunter93} (1993) presented a deep \HA\ imaging and found many
\HA\ filaments around NGC~2363,
though the connection with the high-velocity expanding component is unknown.

In the present observations,
we could detect neither the broad line component
with a few times 1000~km~s$^{-1}$
nor the splitting profile with the expanding velocity of 45~km~s$^{-1}$
at the central region of NGC~2363.
Both of \cite{roy92} (1992) and \cite{gonzalez94} (1994)
extracted the high-velocity expanding
component by picking up the residuals after fitting the emission-line
profile with a single Gaussian and
the peak intensity of the broad component is only 1\% of
that of the observed original line.
We should note that the observed line profile deviates
from the complete Gaussian in general
and that the shape of the wing of the profile depends on the conditions
of the spectrograph;
in our data we could not confirm the existence of the broad line component
with more than 1000~km~s$^{-1}$.
The expanding bubble with the velocity of 45~km~s$^{-1}$ was
clearly shown by \cite{roy91} (1991) in \OIII\ emission line.
The region where the \OIII\ line splits
has an area of about $5'' \times 5''$ as shown
in Fig.~5 of \cite{roy91} (1991).
In our spectroscopy,
one data points in Fig.~2 sampled an area of $1.''8 \times 3.''0$
and the spacing of slit scan was $6.''0$.
Because of the sparse spatial sampling in our scanning,
we may miss the regions showing the line splitting.

\subsubsection{Holmberg II}

The resultant seven position-velocity diagrams are shown in Fig.~3.
The width of the abscissa corresponds to 250$''$ or 3.9~kpc at Holmberg~II.
The observed \HII\ regions are divided into three blocks;
eastern, central, and western blocks.
The eastern block consists of
northern (slit positions A to D) and southern (F and G) parts
and the western block consists of
northern (A and B) and southern (D to F) parts
as shown in the \HA\ map of Fig.~1(b).
Hereafter we call these five regions
Holmberg~II~NE, SE, Center, NW, and SW;
see Fig.~1 (b) and Table~2.

The Center has a steep V-shaped velocity feature
as shown in the diagrams for the slit positions A and B;
the velocity gradient is 30~km~s$^{-1}$ in 200~pc and
the most blue-shifted \HA\ velocity has a smaller velocity
than the \HI\ velocity by about 15~km~s$^{-1}$.
The \HA\ velocity fields for NE, SE, NW, and SW are relatively calm.

\subsubsection{IC 2574}

The resultant six position-velocity diagrams are shown in Fig.~4.
The width of the abscissa corresponds to 200$''$ or 2.9~kpc at IC~2574.
This galaxy has a prominent \HII\ complex in the northeastern part
of the galaxy and this complex consists of several giant \HII\ regions.
Among them the observed positions cover the \HII\ regions
denoted as IC~2574-I and IC~2574-IV in \cite{drissen93} (1993);
see Fig.~1 (c) and Table~2.

The \HA\ velocity crosses the \HI\ velocity upward and downward
on the position-velocity diagrams and
the typical scale of the velocity variation
is 20 -- 30~km~s$^{-1}$ in 200~pc scale,
which is similar to that observed at the outer regions in NGC~2363
as mentioned in Sec.~3.2.1.
Unlike in NGC~2363,
the \HA\ velocity field is chaotic
at whole region of observed \HII\ regions in IC~2574.
Two V-shaped velocity features are prominent in the diagrams
for the slit positions B to D;
the eastern ($170''$ to $140''$)
and the western ($130''$ to $100''$) parts correspond to
the \HII\ regions IC~2574-IV and IC~2574-I,
respectively.
In IC~2574-I,
the velocity gradient at the slit position D is steep,
50~km~s$^{-1}$ in 400~pc,
and the velocity difference between \HA\ and \HI\ at the
most intense \HA\ region is 20~km~s$^{-1}$.
The line width is about 30~km~s$^{-1}$,
a little larger than those in NGC~2366 and Holmberg~II
and rises to about 35~km~s$^{-1}$ at the most \HA-blue-shifted position
in IC~2574-I,
as shown in the diagrams for the slit positions B to D.
\cite{drissen93} (1993) found three candidates of WR stars
in IC~2574-I.
The sharp edge of the velocity feature and
the broad FWHM of the line in IC~2574-I seem to be related to the WR stars.

\subsubsection{WLM}

The resultant ten position-velocity diagrams are shown in Fig.~5.
The width of the abscissa corresponds to 200$''$ or 0.9~kpc at WLM.
As shown in Fig.~1 (d),
the \HII\ regions are in two blocks,
the southern ring-shaped block and the northern barred-shaped block;
hereafter we call them Southern ring and Northern bar,
respectively
(see Fig.~1 (d) and Table~2).

Both blocks of the \HII\ regions are least luminous in the \HA\ light
among 13 observed \HII\ regions.
The velocity fields are flat and
the calmest among four observed galaxies.
The line widths are about 15~km~s$^{-1}$,
which are smaller than those in other three observed galaxies.
The slit positions D to F cross the hole of the Southern ring
and an expanding shell-like feature is seen
at $-100''$ to $-50''$ in the diagram for the slit position E.

\section{Discussion}

\subsection{Characteristics of the H$\alpha$ Position-Velocity Diagrams}

We investigate characteristics of thirteen observed \HII\ regions,
the data of which are summarized in Table~2.
We classify the morphologies of the \HII\ regions into three types;
the first is circular shape filled with \HA\ emission,
the second is ring shape,
and the third is filament.
We call them type-C, -R, and -F morphologies, respectively.
Some \HII\ regions with the type-F morphology seem to be
a chain of small \HII\ regions with the type-C morphology.
The \HA\ luminosities of the \HII\ regions with the type-F morphology
is less than $2 \times 10^{38}$~erg~s$^{-1}$.
We divide the type-C into two subgroups;
type-C1 and C2 for \HA-luminous
($L$(\HA) $\geq 2 \times 10^{38}$~erg~s$^{-1}$) one and \HA-less luminous
($L$(\HA) $< 2 \times 10^{38}$~erg~s$^{-1}$) one,
respectively.

The velocity features in the position-velocity diagrams
are classified into four types and schematically shown in Fig.~6;
we call them type-I, -II, -III, and -IV velocity features, respectively.
The two classifications of the thirteen \HII\ regions are listed
in the last two columns of Table~5.
Table~6 summarizes the classifications of thirteen \HII\ regions
as well as eight \HII\ regions we previously observed in dwarf irregulars
I~Zw~36, Sextans~A, NGC~6822, IC~1613, and NGC~1569
(\cite{tomita93} 1993, 1994).
Table~6 shows that the morphological type and the velocity feature
of the \HII\ regions correlate well with each other;
we characterize the \HII\ regions
into only five categories according to our classifications.

{\it Type-I and type-F or -C2:}~
The position-velocity diagrams for five less luminous \HII\ regions
from NGC~2366 NA to WLM Northern bar in Table~5 are flat.
This is probably due to either lack of energetic stars
which can generate strong wind or
being at a young stage of evolution.
IC~1613 S3,
and probably IC~1613 S2 and Sextans~A
(the \HII\ region at the eastern part of the galaxy)
belong to this category (\cite{tomita93} 1993).

{\it Type-I and type-C1:}~
The velocity fields of powerful \HII\ regions without sharp velocity bumps
suggest that the ages of the \HII\ regions are too young
to generate many evolved stars;
examples are NGC~2366-III, Holmberg~II NE and SE.
At the central part of NGC~2363
the velocity field is flat,
though it is chaotic at outer part.
The velocity field which is calm in the central regions and
chaotic at the outer regions in the luminous \HII\ region
is also observed in NGC~6822 HV (\cite{tomita93} 1993).

{\it Type-II and type-C1:}~
Roy \etal\ (1991) suggested that
the ridge of the \OIII\ velocity field in NGC~2363,
shown as a bump in the position-velocity diagrams
at the positions NGC~2366 IA to ID (see Fig.~2 (a) and 2 (b)),
is related to a chimney.
I~Zw~36,
a well-studied blue compact dwarf galaxy,
has type-II velocity feature (\cite{tomita94} 1994).
\cite{viallefond83} (1983) gave the \HB\ flux of I~Zw~36,
$f$~(\HB) = $3.65 \times 10^{-13}$ erg~s$^{-1}$cm$^{-2}$,
and the distance of 4.6~Mpc.
Assuming $f$(\HA)/$f$(\HB) = 3,
we get $L$~(\HA) = $3 \times 10^{39}$ erg~s$^{-1}$.
The relatively mild velocity field in spite of intense \HA\ luminosity
suggests the \HII\ regions are at a young stage.

{\it Type-III and type-C1:}~
The V-shaped velocity feature associated to two \HII\ regions in IC~2574
may indicate blow-out motions away from the galactic disk.
The \HII\ region HX in NGC~6822 also shows this kind of feature
(\cite{tomita93} 1993).

{\it Type-III or -IV and type-R:}~
The \HA\ ring shows an expanding bubble feature
in the position-velocity diagrams;
we discuss the kinematics
following the analysis given by \cite{mccray87} (1987).
The kinetic age of the bubble is
$t$~[Myr]~= 0.6~$(R/V)$
and the kinetic energy of the bubble is
$E$~[erg]~=  $1.3 \times 10^{42} R^{3} V^{2} n_{0}$,
where $R$ is the radius in kpc,
$V$ is the expanding velocity in km~s$^{-1}$,
and $n_{0}$ is the number density of the ambient interstellar matter
in cm$^{-3}$.
The number of supernova ($N_{\rm SN}$)
responsible for the bubble kinetic energy is derived as
$N_{\rm SN} = E / (0.2 \times 10^{51}$ erg)
assuming that 20\% of the supernova energy of $10^{51}$~erg
contributes to the bubble kinetic energy.
Though we did not observe whole regions of Holmberg~II Center,
we take the expanding velocity for Holmberg~II Center as 30~km~s$^{-1}$
from the data at the position-velocity diagram for the position A.
With a radius of 200~pc,
we get
$t$~=~4~Myr,
$E$~=~$1 \times 10^{52}$~erg,
and $N_{\rm SN}$~=~50,
assuming $n_{0}$~=~1~cm$^{-3}$.
The bubble feature seen in the position NGC~2366 SB
has the same radius and expanding velocity as those for Holmberg~II Center;
200~pc and 30~km~s$^{-1}$, respectively.
\cite{hunter93} (1993) pointed out a possible connection
between the \HA\ filamentary structure around NGC~2363 and
the extraordinary high velocity feature claimed by \cite{roy91} (1991).
The bubble feature seen at NGC~2366 SB corresponds to
a part of the \HA\ filament
(see Fig.~1(a) and mentioned in Sec.3.2.1),
therefore,
this filament was generated by a local bubble,
off the central part of NGC~2363.
WLM Southern ring has
a radius of 150~pc and an expanding velocity of 20~km~s$^{-1}$.
Then,
$t$~=~4.5~Myr,
$E$~=~$2 \times 10^{51}$~erg,
and $N_{\rm SN}$~=~10 are obtained,
assuming $n_{0}$~=~1~cm$^{-3}$.
The age derived is consistent with the result by \cite{ferraro89} (1989)
that the star formation stopped a few Myr ago in the region
including this \HII\ region
(denoted as Region 2 by them).
$N_{\rm SN}$ is similar to those for
expanding shells in NGC~1569 (denoted as SS's in \cite{tomita94} 1994).
The bubbles in our sample have much larger $N_{\rm SN}$ than
those observed by \cite{oey94} (1994) in M~33.

\cite{puche92} (1992) found many \HI\ holes in Holmberg~II.
\cite{puche92} (1992), \cite{mashchenko95} (1995),
and \cite{tongue95} (1995) argued
that they are cavities generated by the star formation activity,
and \cite{hunter93} (1993) claimed a possibility that
the holes were made by the penetration of the high velocity clouds
because the \HA\ emission does not encircle the \HI\ holes
as is the case in LMC
which is expected for cavities by violent star formation.
We do not detect the expanding-bubble velocity features
for the \HII\ regions which surround the \HI\ holes in Holmberg~II
as we observe in WLM.
Our data does not support the idea that the star formation cavities are
responsible for the \HI\ holes in Holmberg~II.

\subsection{Comparison of the Velocity Field between H$\alpha$ and H~I}

The FIR-to-$B$ luminosity ratio,
\LFIRLB,
is an indicator of the present star formation activity of the galaxy
per luminous mass (\cite{tomita96} 1996).
The ratios are 0.3, 0.2, 0.05, and 0.01
for IC~2574, NGC~2366, Holmberg~II, and WLM,
respectively (see Table~1);
the activity ranges over 1.5 order of magnitude.
The ratio of the \HI\ mass to $B$-band luminosity (\MHILB),
on the other hand,
are about 0.1 for all four galaxies (see Table~1).
\cite{huchtmeier81} (1981) presented \HI\ extents
for Holmberg~II, NGC~2366, and WLM.
At a level of $10^{19}$~cm$^{-2}$ of the \HI\ column density,
the radii of the \HI\ extent are 27$'$, 30$'$, and 45$'$,
respectively,
and these correspond to 3.3, 3.6, and 3.8 times the Holmberg radius,
respectively;
the \HI\ to optical size ratios are similar to each other
among the three galaxies.
The average \HI\ surface densities
are several to ten times $10^{5}$~\Msolar~kpc$^{-2}$ and
are also similar to each other among the three galaxies.
No correlation between the star formation activity,
such as \LFIRLB,
and the \HI\ density
suggests that some external causes or internal but intermittent causes
trigger the generation of the star forming regions
as well as cloud formation by the self gravity or
shell sweeping (e.g., \cite{oey95} 1995).

\cite{saito92} (1992) proposed a model
based on observed \HA\ and \HI\ velocity fields of IC~10
that infalling clouds from the extended \HI\ envelope of the galaxy
may cause the intense star formation.
In their model,
the \HII\ regions made by infalling \HI\ clouds have bulk motions of about
10~km~s$^{-1}$ relative to the ambient \HI\ disk.
The chaotic velocity fields,
with bubbles or blow-out features through the intense star formation,
presented by types-III and IV,
would disturb the information about the original bulk motion
of star-forming clouds.
On the other hand,
the calm \HA\ velocity field,
types-I and II,
is suitable
to measure the original bulk motion of the \HII\ regions.
We investigate eight \HII\ regions;
NGC~2363 (the slit position of NGC~2366 SG),
NGC~2366-III (the slit position of NGC~2366 SD),
NGC~2366 NA,
NGC~2366 ND,
Holmberg~II NE,
SE, NW, and SW.

Table~7 lists the \HI\ and \HA\ radial velocities
as well as the velocity difference between them.
In three out of eight,
NGC~2363, Holmberg~II NE, and Holmberg~II NW,
we detect a little or no velocity difference.
In other three \HII\ regions in NGC~2366,
we detect velocity differences of about 10~km~s$^{-1}$.
We detect velocity differences of 15 to 20~km~s$^{-1}$
for other two \HII\ regions in Holmberg~II.
The distribution of the velocity differences among the eight \HII\ regions
ranges $-$10~km~s$^{-1}$ to 20~km~s$^{-1}$
and the average is several km~s$^{-1}$.
In the previous study by \cite{tomita93} (1993),
five \HII\ regions,
I~Zw~36, Sextans~A, IC~1613 S2 and S3, and NGC~6822 HV,
in four galaxies have type-I or -II velocity features (see Table~6).
Fig.~7 summarize in histogram the observed
velocity differences of \HA\ $-$ \HI\
for \HII\ regions with type-I or -II velocity features
including above five \HII\ regions.
The standard deviation of the velocity difference is about 5~km~s$^{-1}$,
which is consistent with the expectation from the model by
\cite{saito92} (1992).
Though it is marginal and it is not conclusive
because such a velocity difference may be generated by
internal motion through star formation activity,
this suggests that some of the \HII\ regions in dwarf irregulars
formed through the infall of clouds onto the galactic disk.
Search for high velocity clouds around the galaxies
are needed to confirm this hypothesis.

\acknowledgments

We would like to thank Yoh-ichi Kanamori
for his help with the observations,
and the OAO stuff members
for their hospitality during our stay.
T.T.T. acknowledges the Research Fellowship
of the Japan Society for the Promotion of Science for
Young Scientists.
Finally,
we are grateful to the anonymous referee for improving the paper.

\clearpage

\figcaption[fig1a.ps, fig1b.ps, fig1c.ps, fig1d.ps]{
The slit positions superimposed on the \HA\ images.
The length of each line corresponds to the length of the abscissa
of Fig.~2.
(a) NGC~2366:
the \HA\ image is taken from Roy \etal\ (1991)'s Fig.~1.
The length of the line corresponds to 160$''$.
(b) Holmberg~II:
the \HA\ image is taken from Hodge \etal\ (1994)'s Fig.~1.
The length of the line corresponds to 250$''$.
(c) IC~2574:
the \HA\ image is taken from Miller \& Hodge (1994)'s Fig.~3.
The length of the line corresponds to 200$''$.
(d) WLM:
the \HA\ image is taken from Hodge \& Miller (1995)'s Fig.~1 (a).
The length of the line corresponds to 200$''$.
\label{fig1}}

\figcaption[fig2a.ps, fig2b.ps, fig2c.ps, fig2d.ps, fig2e.ps,
fig2f.ps, fig2g.ps, fig2h.ps]{
Position-velocity diagrams for NGC~2366
(Fig.~2 (a) -- 2 (h)).
Sixteen panels are presented and the slit position name
is labeled at the top of each panel.
The abscissa indicates the position along the slit in arcsec
and the coordinates of the origin is listed in Table~4.
The ordinate indicates the heliocentric radial velocity
and the FWHM of the \HA\ emission line.
A filled circle indicates the central velocity of the line
and an open circle indicates the FWHM of the line.
The scales for the central velocity and the FWHM of the line
are shown in the left-side and the right-side ordinates,
respectively.
The horizontal and vertical error bars on a filled circle show
the binning width and the error of measurement of the central velocity,
respectively.
The instrumental broadening is reduced from the observed line width
to derive the FWHM.
The dotted line shows the 0~km~s$^{-1}$-level for the FWHM.
The thick curve indicates the H~I velocity.
The data is taken from Wevers \etal\ (1986).
The cross in the upper-left corner of the panel
shows the accuracy of the H~I observations.
The full length of the horizontal line of the cross,
$24''$,
indicates the FWHM of the beam
for the H~I isovelocity map.
For four positions of NGC~2366~IA to ID,
the FWHM is 25$''$,
since the size of the beam depends on the position angle.
The half length of the vertical line of the cross,
8.25~km~s$^{-1}$,
indicates the channel spacing of the H~I observations.
\label{fig2}}

\figcaption[fig3a-d.ps]{
The same as Fig.~2 but for Holmberg~II.
The H~I data is taken from Puche \etal\ (1992).
The beam size and the channel spacing of the H~I observations are
4.5$''$ and 2.58~km~s$^{-1}$,
respectively.
\label{fig3}}

\figcaption[fig4a-c.ps]{
The same as FIg.~2 but for IC~2574.
The H~I data is taken from Martimbeau \etal\ (1994).
The beam size and the channel spacing of the H~I observations are
30$''$ and 8.24~km~s$^{-1}$,
respectively.
\label{fig4}}

\figcaption[fig5a-e.ps]{
The same as Fig.~2 but for WLM.
There is no synthetic H~I observations in literature so far.
\label{fig5}}

\figcaption[fig6.ps]{
Schematic pictures of four types of the velocity features
in the position-velocity diagram.
In ordinate,
the upper direction indicates larger recession velocity.
The type is shown in the upper right corner in each panel.
\label{fig6}}

\figcaption[fig7.ps]{
Histogram of the bulk motion of the H~II regions
relative to the galactic H~I disk defined by
\HA\ $-$ H~I.
The data for thirteen H~II regions with type-I or -II velocity feature
including the samples observed by Tomita \etal\ (1993) are presented.
\label{fig7}}

\end{document}


\newcommand {\etal}{{\it et al.}}
\newcommand {\HI}{H~{\small I}}
\newcommand {\HII}{H~{\small II}}
\newcommand {\HA}{H$\alpha$}
\newcommand {\HB}{H$\beta$}
\newcommand {\NII}{[N~{\small II}]}
\newcommand {\OIII}{[O~{\small III}]}
\newcommand {\degree}{$^{\circ}$}
\newcommand {\Msolar}{$M_{\odot}$}
\newcommand {\Lsolar}{$L_{\odot}$}
\newcommand {\LFIRLB}{$L_{FIR}/L_{B}$}
\newcommand {\fFIRfB}{$f_{FIR}/f_{B}$}
\newcommand {\MHILB}{$M_{\rm H I}/L_{B}$}

\begin{table}[h]

\begin{center}


T{\small ABLE} 2: Characteristics of the observed H~II regions.

 \begin{tabular}{llrrcrrll}
 \tableline \tableline
 Galaxy      & H~II region  & Other ID  &$f$(\HA)&Mor. &
                                            $\Sigma~f$(\HA)&$L$(\HA)&Ref&Note\\
 ~~~~(1)     & ~~~~~(2)      & (3)~~~~   & (4)~~~~& (5) &(6)~~&(7)~~&   &    \\
 \tableline
 NGC 2366    & NGC 2363  & NGC 2366-I,II &6710~~~~& b   &6710 &95~~ &1,2& (i)\\
 \cline{2-9}
             & NGC 2366-III  &           &1000~~~~& b   &1000 &14~~ &   &(ii)\\
 \cline{2-9}
             & NGC 2366 NA   &           & 100~~~~& b   & 100 & 1.4 &  &(iii)\\
 \cline{2-9}
             & NGC 2366 ND   &           & 100~~~~& b   & 100 & 1.4 &  &(iii)\\
 \tableline
 Holmberg II & NE            & Hodge~~71 & 159~~~~& b   & 593 & 7.3 & 3 &    \\
             &               &        73 & 339~~~~& l   &     &     &   &    \\
             &               &        74 &  95~~~~& b   &     &     &   &    \\
 \cline{2-9}
             & SE            & Hodge~~70 & 244~~~~& c   & 244 & 3.0 & 3 &    \\
 \cline{2-9}
             & Center        & Hodge~~45 & 754~~~~&b,c,l& 754 & 9.2 & 3 &    \\
 \cline{2-9}
             & NW            & Hodge~~16 & 42.6~~ & b   & 103 & 1.3 & 3 &    \\
             &               &        17 & 60.3~~ & b   &     &     &   &    \\
 \cline{2-9}
             & SW            & Hodge~~18 & 19.0~~ & f   &  90 & 1.1 & 3 &    \\
             &               &        20 & 71.1~~ & b   &     &     &   &    \\
 \tableline
 IC 2574     & IC 2574-I     & Hodge 167 & 502.93 & b   &1095 &12~~ & 4 &(iv)\\
             &               &       169 & 313.54 & b   &     &     &   &    \\
             &               &       179 &  79.29 & b   &     &     &   &    \\
             &               &       180 & 108.99 & b   &     &     &   &    \\
             &               &       185 &  90.43 & b   &     &     &   &    \\
 \cline{2-9}
             & IC 2574-IV    & Hodge 194 & 191.08 & b   & 487 & 5.3 & 4 &(iv)\\
             &               &       198 & 167.04 & b   &     &     &   &    \\
             &               &       202 & 129.13 & b   &     &     &   &    \\
 \tableline
 WLM         & Northern bar  &Hodge~~~~2 &  52.74 & b   & 318 & 0.4 & 5 &    \\
             &               &         3 &  13.68 & f   &     &     &   &    \\
             &               &         4 &  40.72 & b   &     &     &   &    \\
             &               &         5 &   9.22 & f   &     &     &   &    \\
             &               &         8 &  54.44 & b   &     &     &   &    \\
             &               &         9 & 107.79 & b   &     &     &   &    \\
             &               &        13 &   9.44 & f   &     &     &   &    \\
             &               &        14 &  10.29 & f   &     &     &   &    \\
             &               &        15 &  19.88 & f   &     &     &   &    \\
 \cline{2-9}
             & Southern ring &Hodge~~~~6 &  45.54 & b   & 299 & 0.3 & 5 &    \\
             &               &         7 & 159.23 & b   &     &     &   &    \\
             &               &        10 &  10.86 & f   &     &     &   &    \\
             &               &        11 &  37.54 & b   &     &     &   &    \\
             &               &        12 &  45.86 & f   &     &     &   &    \\
 \tableline \tableline
 \end{tabular}

\end{center}

\rightskip=1cm
\leftskip=1cm
\bigskip

Columns:

\noindent
(1) Galaxy name.
(2) Name of the H~II regions in this paper.
(3) Name of individual H~II regions in other papers
mainly by Hodge and his collaborators which are here denoted as `Hodge \#'.
(4) Flux of the \HA;
the reference is shown in the eighth column.
The unit is $10^{-15}~$erg~s$^{-1}$cm$^{-2}$.
(5) Hodge-type morphology;
c: compact, b: bright compact,
f: faint compact, l: loops or rings, and d: diffuse emission.
(6) Total flux of the \HA\ summing up the values in the fourth column.
The unit is $10^{-15}~$erg~s$^{-1}$cm$^{-2}$.
(7) Luminosity  of the \HA\
in the unit of $10^{38}$~erg~s$^{-1}$.

\bigskip
References:

\noindent
(1) Mayya (1994);
(2) Kennicutt \etal\ (1980);
(3) Hodge \etal\ (1994);
(4) Miller \& Hodge (1994);
(5) Hodge \& Miller (1995)

\end{table}

\clearpage

\rightskip=2cm

{\centerline {NOTES:}}

(i)
We combine NGC~2366-I and II as NGC~2363
because of close position,
though originally NGC~2363 corresponds to NGC~2366-I.
We derive the \HA\ flux using values
log~$f$(\HA\ + \NII) [erg~s$^{-1}$cm$^{-2}$] = $-$ 11.16
given by Mayya (1994) for the H~II region with ID `1~+~2',
and \NII/\HA\ $\sim$ 0.03 given by Kennicutt \etal\ (1980).
We give the morphology following Hodge's classification.
Note that NGC~2363 can be divided into many small H~II regions.

(ii)
We estimate the \HA\ flux considering that the surface brightness
and area are 0.6 and 0.25 of those for NGC~2363,
respectively.
We give the morphology following Hodge's classification.
Note that NGC~2366-III can be divided into many small H~II regions.

(iii)
We estimate the \HA\ flux considering that the surface brightness
and area are 0.25 and 1/15 of those for NGC~2363,
respectively.
We give the morphology following Hodge's classification.

(iv)
Martimbeau \etal\ (1994) reported that
$f$(\HA) [erg~s$^{-1}$cm$^{-2}$] = $7.8 \times 10^{-13}$ and
$4.7 \times 10^{-13}$ for IC~2574-I and IV,
respectively.